\documentclass{tMPH2e}

\newcommand\acie{Angew. Chem. Int. Ed.\ }
\newcommand\cpc{Comp. Phys. Comm.\ }
\newcommand\jacs{J. Am. Chem. Soc.\ }
\newcommand\jcp{J. Chem. Phys.\ }
\newcommand\jmst{J. Mol. Struct. (Theochem.)\ }
\newcommand\jpcb{J. Phys. Chem. B\ }
\newcommand\jpcm{J. Phys. Cond. Mat.\ }
\newcommand\molp{Mol. Phys.\ }
\newcommand\ms{Mol. Simul.\ }
\newcommand\nat{Nature\ }
\newcommand\pmag{Philos. Mag.\ }
\newcommand\pnasu{Proc. Natl. Acad. Sci. USA\ }
\newcommand\pmsix{Philos. Mag. Ser. VI\ }
\newcommand\pr{Phys. Rev.\ }
\newcommand\prb{Phys. Rev. B\ }
\newcommand\sm{Soft Matter\ }

\begin{document}

\doi{10.1080/0026897YYxxxxxxxx}
 \issn{1362 3028}
\issnp{0026 8976}
\jvol{00}
\jnum{00} \jyear{2015} 

\title{Dynamics and thermodynamics of decay in charged clusters}

\author{Mark A.~Miller$^{a}$$^{\ast}$\thanks{$^\ast$Corresponding author. Email: m.a.miller@durham.ac.uk
\vspace{6pt}}, David A.~Bonhommeau$^{b}$, Christian P.~Moerland$^{c}$, Sarah J.~Gray$^{a}$
and Marie-Pierre Gaigeot$^{d}$\\\vspace{6pt}
$^{a}${\em{Department of Chemistry, Durham University, South Road, Durham, UK}}\\
$^{b}${\em{GSMA CNRS UMR7331, Universit{\'e} de Reims Champagne-Ardenne,
UFR Sciences Exactes et Naturelles, Moulin de la Housse, BP 1039, F-51687 Reims Cedex 2, France}}\\
$^{c}${\em{Department of Applied Physics, Technische Universiteit Eindhoven,\\
Postbus 513, 5600 MB Eindhoven, The Netherlands}}\\
$^{d}${\em{Universit{\'e} d'Evry val d'Essonne, LAMBE CNRS UMR8587,
Boulevard F Mitterrand, B{\^a}timent Maupertuis, F-91025 Evry, France}}
}

\maketitle

\begin{abstract}
We propose a method for quantifying charge-driven instabilities in clusters,
based on equilibrium simulations under confinement at constant external pressure.
This approach makes no assumptions about the mode of decay and allows different
clusters to be compared on an equal footing.  A comprehensive survey of stability
in model clusters of 309 Lennard-Jones particles augmented with Coulomb interactions
is presented.  We proceed to examine dynamic signatures
of instability, finding that rate constants for ejection of charged particles
increase smoothly as a function of total charge with no sudden changes.  For clusters
where many particles carry charge, ejection of individual charges competes with a fission
process that leads to more symmetric division of the cluster into large fragments.
The rate constants for fission depend much more sensitively on total charge than those
for ejection of individual particles.

\bigskip

\end{abstract}

\section{Introduction}

When a droplet or cluster carries a sufficiently high net electrostatic charge, the
charge drives decay into two or more fragments.  The first
analysis of this phenomenon, due to Rayleigh \cite{Rayleigh82a} is now well over
a century old, but the resulting prediction for the critical charge at which
decay occurs retains its central importance in modern research in the area.
Rayleigh's analysis is based on a uniformly charged, structureless
sphere of diameter $D$ that may undergo shape fluctuations, leading simultaneously
to an increase
in surface energy and a decrease in electrostatic energy.  Deformation leads to a
barrierless decrease in the sum of these energies beyond the limiting charge
$Q_{\rm R}=\pi\sqrt{8\epsilon_0\gamma D^3}$, where $\gamma$ is the surface tension
of the droplet interface and $\epsilon_0$ is the permittivity of free space.
Rayleigh reached this classic result within the first page of his elegantly short
article \cite{Rayleigh82a}, concealing a somewhat involved derivation \cite{Consta15a}.
\par
Since Rayleigh's time, his work has gained new significance due to the importance
of charged droplets in contemporary fields such as atmospheric chemistry \cite{Hirsikko11a}
and electrospray ionisation mass spectrometry \cite{Fenn03a}.  Nevertheless,
it has been recognised that unstable global deformations are not the only
mechanism by which a charged droplet may decay.  For example, the ion
evaporation model \cite{Iribarne76a} provides a continuum-based
understanding of how small ions may be ejected from a droplet even below the Rayleigh
limit.  When macroions are dissolved in a droplet, other mechanisms come into play.
Evaporation of the droplet may lead to some of the charge being deposited on
the solute \cite{Dole68a}.
Unravelling of polymeric solutes in water can lead to more complex
processes \cite{Consta10a,Ahadi12a},
including the extrusion of a solvent-free segment of the chain or the formation
of subdroplets along extended chain conformations \cite{Consta13a}.
\par
In principle, a cluster at finite temperature in an unbounded environment is
thermodynamically at most metastable, even if it is charge-neutral, since
unlimited translational entropy can
be gained by dividing the cluster into fragments.  Hence, the concept of stability
implicitly involves a sense of time scale, such as the typical time of flight of
an electrosprayed droplet in a mass spectrometer.  In a previous paper, some
of us devised an equilibrium approach to identifying charge-driven instability in
small clusters \cite{Miller12a} based on confining the cluster to a spherical
container and monitoring the fraction of equilibrium configurations in which the
cluster had emitted at least some of its charged particles.  In the present
article, we propose a different equilibrium-based approach that provides some significant
advantages in terms of the generality of its applicability and the physical
basis of its interpretation.  We then proceed to a quantitative characterisation
of the dynamics of charge-driven
instabilities in a simple model, focusing on the competition between ejection
and fission processes.

\section{Model}

As in our previous study \cite{Miller12a}, we use a Lennard-Jones (LJ) cluster
of $N=309$ particles to explore methods of characterisation.  For simplicity,
all particles interact with the same LJ well-depth $u$ and length
parameter $\sigma$.  In addition, particles may carry a charge, introducing
a Coulomb term to the total potential energy:
\begin{displaymath}
E=4u\sum_{i<j}^N\left[\left(\frac{\sigma}{r_{ij}}\right)^{12}
-\left(\frac{\sigma}{r_{ij}}\right)^6\right]+
\sum_{i<j}^N\frac{q_i q_j}{4\pi\epsilon_0 r_{ij}},
\end{displaymath}
where $q_i$ is the charge on particle $i$ and $r_{ij}$ is the distance between
particles $i$ and $j$.  We adopt the usual LJ reduced energy $E^*=E/u$,
length $r^*=r/\sigma$ and temperature $T^*=k_{\rm B}T/u$, where $k_{\rm B}$ is
Boltzmann's constant.  We also define a dimensionless charge
$q^*=q_i/\sqrt{4\pi\epsilon_0\sigma u}$.  In terms of reduced variables, therefore,
\begin{displaymath}
E^*=\sum_{i<j}^N\left[4({r_{ij}^*}^{-12}-{r_{ij}^*}^{-6})+q_i^* q_j^*/r_{ij}^*\right].
\end{displaymath}
Since the reduced charge $q^*$ depends on $u$ and $\sigma$ in addition to $q$, it is not confined
to integer multiples of the electronic charge.  To enable us to examine properties
of the model charged cluster systematically
as a function of charge, we will treat the reduced charge
as a continuous variable.  However, we will restrict ourselves to the case where
$n$ of the $N$ particles carry identical single-particle charges, $q^*_i=q^*$, while
the remaining $N-n$ particles are neutral, with $q_i^*=0$.  The total reduced
charge on the cluster is then $Q^*=nq^*$.
\par
$N=309$ is a magic number for the LJ potential, corresponding to four
complete icosahedral shells in the lowest-energy structure \cite{Romero99a}.
The neutral cluster's thermodynamics has been studied in detail \cite{Noya06a},
allowing us to select a reduced temperature of $T^*=0.43$, which lies just
above the broadened melting transition, for most of our calculations.
In this liquid-like regime, the highly ordered global potential minimum
structure is not seen.  As might be expected, the addition of charge lowers
the cluster's melting temperature \cite{Bonhommeau13a}.
\par
The LJ potential with charges localised on individual particles is
clearly a highly idealised model.  It does not incorporate the complexity
associated with hydrogen bonding in water clusters or mechanisms for transferring
excess charge to the cluster surface by orientation of dipole moments \cite{Vacha11a}.
We note that it is possible to refine the model by the inclusion of polarisability,
with induced dipoles solved self-consistently \cite{Bonhommeau13a}, though this is
a costly addition to a simple model.  Nevertheless, idealised models of this type
do provide scope for probing different mechanisms of decay \cite{Last02a,Levy06a},
and can even provide a useful reference for quantitative analysis of experimental data
on the evaporation of real charged droplets \cite{Holyst13a}.

\section{Droplets at equilibrium}

As pointed out in the introduction, an unbounded cluster decays over time even if
it is uncharged.  Clusters bound by van der Waals-like interactions such as the LJ
potential gradually evaporate, typically one particle at a time.  Hence, the
equilibrium thermodynamics of clusters must always be characterised within a
container.  A spherical enclosure is usually used, and an appropriate radius must then be
chosen by some criterion \cite{Lee73a,Tsai93a}.  In our previous article \cite{Miller12a},
we used a container radius of $R_{\rm c}=6.5\sigma$ for the 309-atom LJ cluster, with
the centre of the container tracking the centre-of-mass of the cluster \cite{Lee73a}.
This choice of $R_{\rm c}$ allows a clear separation between any particles that have
been decisively ejected and the remaining subcluster.  However, the equilibrium between
evaporated and condensed particles is influenced by the choice of radius \cite{Tsai93a}
and so the arbitrary choice of $R_{\rm c}$ is an unattractive feature of the approach.
It is also difficult to compare clusters of different size $N$, since it is not
entirely clear how $R_{\rm c}$ should change with $N$ for a truly unbiased comparison.
\par
In the present work, we take a different approach that offers a number of advantages.
We place the cluster in a fluctuating container whose volume changes
in response to the cluster inside it
and to a fixed external pressure $p$.  This pressure may be set at a physically
meaningful value by reference to, for example, the background medium or carrier
gas in a particular application.  The same value of $p$
may straightforwardly be used for different $N$ if a comparison between cluster
sizes is required.
\par
The isobaric, isothermal approach that we propose is slightly different from
typical constant-$NpT$ Monte Carlo simulations, which are normally encountered
when modelling bulk systems as represented by a periodic simulation cell.
In a periodic system, stochastic changes in the cell volume are accompanied by a uniform
scaling of all particle positions in order to respect the boundary conditions in the
resized cell, and particle displacement steps are made in coordinates that have been
scaled to lie in a unit cube.
This protocol leads to a Monte Carlo acceptance criterion for trial volume changes
of \cite{Frenkel02b}
\begin{displaymath}
P^{\rm acc} = {\rm min}\left[1,\ \exp\left(-\frac{(E'-E)}{k_{\rm B}T}\right)
\exp\left(-\frac{p(V'-V)}{k_{\rm B}T}\right)
\left(\frac{V'}{V}\right)^N\right],
\end{displaymath}
where $V$ is the volume of the cell and primes indicate new values after the trial move.
The factor containing $E$ is due to the energy change associated with scaling the particle
coordinates, while the factor $(V'/V)^N$ comes from the change of variables to scaled
coordinates in a unit cell.  However, our cluster is not a periodic system, and it is more
efficient to allow the container to fluctuate without unnaturally expanding or contracting
the cluster by a scaling of the particle coordinates.
If the configuration is left unchanged during trial volume changes then
the acceptance for trial volume changes is the simpler
\begin{displaymath}
P^{\rm acc} = {\rm min}\left[1,\ \exp\left(-\frac{p(V'-V)}{k_{\rm B}T}\right)
\prod_{i=1}^N \Theta(R'_{\rm c}-|{\bf r}'_i|)\right],
\end{displaymath}
where ${\bf r}'_i$ is the position of particle $i$ relative to the centre of mass (after the trial move).
The product of Heaviside step functions $\Theta$ enforces the requirement that all particles
lie within the container.  Hence, any trial decrease in volume that would result in the exclusion
of an outlying particle must be rejected.  In effect, such a proposed state would have
infinite energy because the container is hard.  Trial volume changes constitute 1\%
of Monte Carlo steps in our simulations.
\par
In addition to the modified volume-change steps, 20\% of Monte Carlo steps in our
simulation are devoted to attempted exchanges between neutral and charged particles, leaving
their positions fixed.  These swaps are accepted according to the standard Metropolis criterion
\cite{Metropolis53a,Frenkel02b} and make exploration of the equilibrium configurations more
efficient \cite{Miller12a}.

\begin{figure}
\begin{center}
\includegraphics[width=80mm]{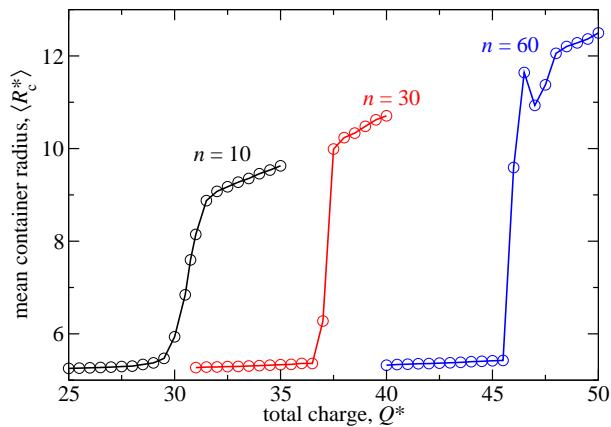}
\end{center}
\caption{Detection of decay using the fluctuating-container method at reduced pressure $p^*=0.005$
for reduced total charge $Q^*$ distributed amongst three different numbers $n$ of charge-carriers.
Instability appears as a sharp increase in the reduced mean container radius.
\label{jumps}
}
\end{figure}

Figure \ref{jumps} shows the mean reduced container radius $\langle R^*_{\rm c}\rangle$
as a function of total charge $Q^*$ at a reduced pressure of $p^*=p\sigma^3/u=0.005$.
(For reference, with argon LJ parameters, 1 bar corresponds to $p^*\approx0.0024$.)
Each point in the figure was obtained from a simulation of $2\times10^7$
Monte Carlo steps per particle. A sharp
increase in $\langle R^*_{\rm c}\rangle$ occurs at a particular charge, indicating that
the fragmentation of the cluster is sufficient to force the container outwards against
the external pressure.  From a thermodynamic point of view, the sudden change is
akin to the finite-system analogue of a first-order liquid--gas phase
transition, but here driven by charge rather than by temperature.  The value of
$Q^*$ at which the jump occurs increases with the
number of particles $n$ over which the charge is divided, reinforcing our earlier
result that the maximum charge a cluster can sustain is generally larger when the charge
is divided over more particles \cite{Miller12a}.

\begin{figure}
\begin{center}
\includegraphics[width=80mm]{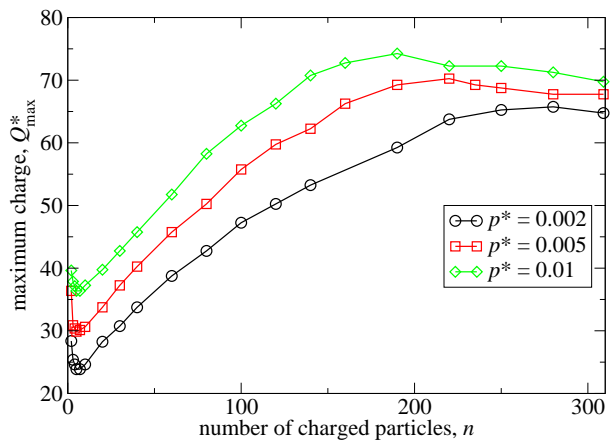}
\end{center}
\caption{Maximum charge as detected by the fluctuating-container method at three reduced pressures
$p^*$.
\label{Qmax}
}
\end{figure}

We take the point at which the numerical derivative $dR_{\rm c}^*/dQ^*$ is largest
to be the maximum charge $Q^*_{\rm max}$ that the cluster can sustain.  This transition is
more sharply defined than the point at which dissociated configurations become a majority
in a container of fixed radius \cite{Miller12a}.  $Q^*_{\rm max}$ based on the
fluctuating-container method is plotted as a function of $n$ in Fig.~\ref{Qmax},
showing qualitatively similar behaviour at three different pressures.  At large $n$,
$Q^*_{\rm max}$ shows a small but systematic decrease.  As will be explored in the next
section, this feature corresponds to the onset of competition between ejection of
individual charged particles and fission of the cluster into larger fragments.
The fact that the fluctuating container detects both these modes of decay using the same
procedure and without prior knowledge of the change in mechanism is one of the method's
appealing features.  Even in the more complex cases of droplets of polar molecules
with or without a polymeric solute, the decay pathways all involve spreading out of the
charged particles by a decisive deviation of the droplet away from a spherical
or near-spherical shape \cite{Consta10a,Ahadi12a}.  Such changes would always lead to
a sudden increase in the radius of the container at some critical charge.  Hence,
the fluctuating-container method should be quite generally applicable to the problem
of detecting and defining charge-driven instability.
\par
At very small $n$, Fig.~(\ref{Qmax}) shows that there is a shallow
minimum in $Q^*_{\rm max}$.  This feature seems to be related
to the rapid increase in energy with respect to the number of charges at small $n$.  In the
hypothetical case where the charged particles lay neatly at the surface of a
perfectly spherical cluster,
the cluster would be described by the so-called Thomson problem \cite{Thomson04a}, which poses
the question of the optimal arrangement and corresponding energy of $n$ unit charges on the surface
of a unit sphere.  In the continuum limit, a sphere with a
uniformly charged surface has an electrostatic energy that scales as the square of the
surface charge density.  The global potential energy minima of the Thomson problem do
approach a constant ratio of energy to $n^2$ at large $n$ \cite{Wales06b}.  However, for
small $n$, the energy rises considerably more steeply than $n^2$ \cite{Munera86a}.

\section{Dynamics of decay}

We now turn to dynamic signatures of charge-driven instability.  The clusters must be
prepared in a well defined intact state before being allowed to decay.  Our protocol is to
equilibrate the cluster using constant-temperature Monte Carlo simulations
in a tight-fitting spherical container of fixed radius $5.2488\sigma$, which is
the maximum radius of the icosahedral global minimum structure \cite{Romero99a}
plus a margin of $\sigma$.  The container is then removed and random
Maxwell--Boltzmann-distributed velocities are assigned to the atoms.  From this point,
the cluster is allowed to evolve via constant-energy molecular dynamics (MD)
with a standard Verlet integrator \cite{Frenkel02b} and a time-step of
$\delta t^*=0.002$, where the LJ reduced time is
$t^*=t\sqrt{u/m\sigma^2}$ and $m$ is the mass of one particle.
\par
To detect evaporation and ejection events, we monitor the distance of all particles from
the centre of mass.  When a particle reaches a large distance ($22\sigma$) from the
remaining subcluster, we can be confident that it will not return.  The actual time of the
evaporation or ejection is then backdated to the MD step at which the velocity of the particle
first started on its outward trajectory, {\em i.e.}, the earliest time $t_{\rm e}^*$ such
that ${\bf v}_i(t^*) \cdot {\bf r}_i(t^*) > 0$ for all $t^*\ge t_{\rm e}^*$, where
${\bf v}_i$ is the velocity of particle $i$.  This procedure allows temporary
excursions of particles to be ignored while still providing an accurate timestamp for
ejection events.

\begin{figure}
\begin{center}
\includegraphics[width=80mm]{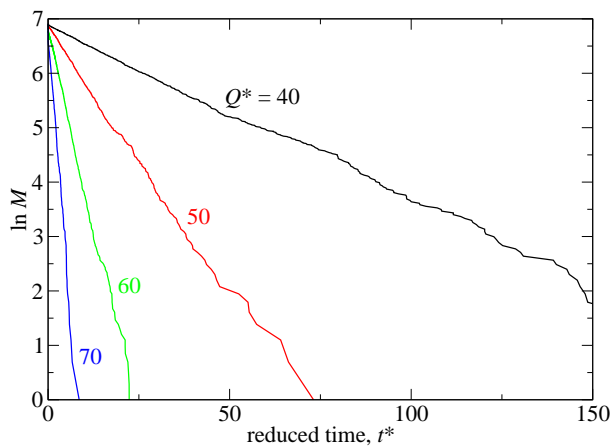}
\end{center}
\caption{First-order decay plots for first ejection of a charged particle averaged
over an ensemble of 1000
clusters carrying $n=100$ charges at four values of the initial total charge $Q^*$.
$M$ is the number of intact clusters remaining at time $t^*$.
\label{FirstOrder}
}
\end{figure}

\begin{figure}
\begin{center}
\includegraphics[width=80mm]{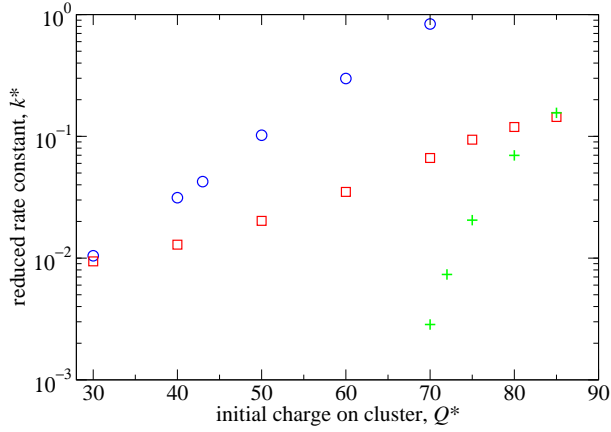}
\end{center}
\caption{Rate constants for the ejection of the first particle for a cluster carrying $n=100$ charged
particles (blue circles) and $n=309$ charged particles (red squares), and rate constants for
fission of the $n=309$ cluster (green plusses).
\label{RateConstants}
}
\end{figure}

We have measured the time of first ejection of a charged particle in an ensemble of
1000 independent trajectories.  From the list of ejection times, we may construct a
first-order kinetics plot of the logarithm of the number $M$ of clusters that 
have not yet decayed as a function of time.  Results for a selection of ensembles
with $n=100$ charged particles and different initial total charges are shown
in Fig.~\ref{FirstOrder}, confirming that ejection of the first charge is a first-order
process.  From the slopes, we obtain reduced rate constants $k^*$, which are plotted
(circles) as a function of initial charge in Fig.~\ref{RateConstants}.

The total charge spanned in Fig.~\ref{RateConstants} starts well below and finishes
considerably above the range of $Q^*_{\rm max}$ measured in our equilibrium simulations
(Fig.~\ref{Qmax}) for $n=100$.  The rate constant increases smoothly with $Q^*$, never
showing a decisive jump that could be taken as a dynamically-defined threshold for instability.

\begin{figure}
\begin{center}
\includegraphics[width=80mm]{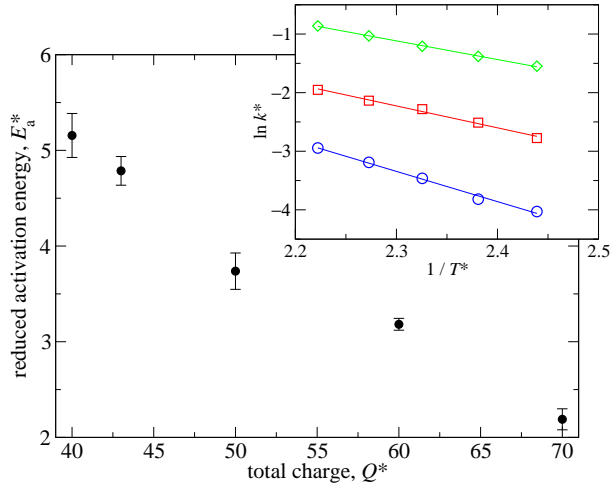}
\end{center}
\caption{Activation energy $E^*_{\rm a}$ for particle ejection for a cluster carrying $n=100$
charges.  Bars show the standard error in the linear regression coefficient.
The inset shows example Arrhenius plots whose slopes are $-E^*_{\rm a}$: $Q^*=40$
(blue circles), $Q^*=50$ (red squares), $Q^*=60$ (green diamonds).
\label{Arrhenius}
}
\end{figure}

We may also extract an effective activation energy $E^*_{\rm a}$ for ejection from
Arrhenius plots of $\ln k^*$ against inverse temperature.  Remembering the proximity
of our chosen temperature $T^*=0.43$ to the melting transition of the neutral cluster
\cite{Noya06a} on the one hand, and the rapidly increasing tendency for thermal
evaporation at higher temperatures \cite{Tsai93a} on the other, we have varied $1/T^*$
over only a narrow range.  Arrhenius behaviour is nevertheless clearly evident, as
shown in the inset of Fig.~\ref{Arrhenius}, allowing the activation energy to be
plotted as a function of total charge in the main panel of the figure.  In keeping
with the smoothly changing rate constants of Fig.~\ref{RateConstants}, the activation
energy shows no sudden changes with charge and appears to depend rather linearly on
charge in this case.

In highly charged clusters, the ejection of a charged particle can lower the overall
potential energy of the remaining subcluster and, depending on the kinetic energy carried
away, can in turn lead to slight increase in the effective temperature of the subcluster.
However, we have not observed any ``run-away'' cascades where the rate of decay of an
ensemble increased with successive ejection events.

The same observation of smoothly changing first-ejection rates at $n=100$ applies to the case of
$n=309$ (squares in Fig.~\ref{RateConstants}), where the charge is spread over all the
particles in the cluster.  As might be predicted from the higher value of $Q^*_{\rm max}$
at large $n$ (Fig.~\ref{Qmax}), the rate constant is lower than for the $n=100$ case at a
given total charge, despite the fact that there are more charged particles that could be
ejected at $n=309$.

\begin{figure}
\begin{center}
\includegraphics[width=80mm]{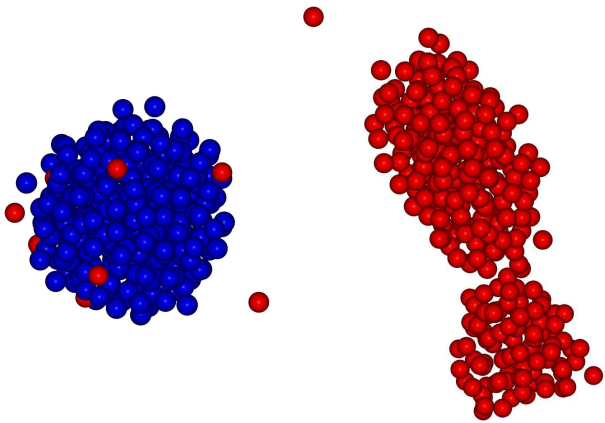}
\end{center}
\caption{Left: ejection of a particle from a cluster carrying $n=10$ charges (red spheres),
total charge $Q^*=24$.
Right: fission of a cluster of $n=309$ charges, $Q^*=67$, into two sub-clusters.
\label{snapshots}
}
\end{figure}

The crucial difference between $n=100$ and $n=309$ is that ejection of individual charged
particles is not the only mechanism of decay for the latter.  At large $n$,
clusters may also undergo a more symmetric
fission process, in which two large fragments are produced.  Snapshots of ejection and
fission processes are shown in Fig.~\ref{snapshots}.  Fission may be detected by performing
a cluster analysis based on a pairwise particle connectivity criterion of $1.5\sigma$ and
defining fission to have taken place at the first time at which two separate
clusters, each of at least
$N/4$ particles, exist.  (The results are insensitive to the arbitrary choices of connectivity
criterion and subcluster size.) Since the clusters are equilibrated
in a tightly-fitting spherical container, the elongation fluctuations that lead to
fission take some time to be established before fission itself can take place.  Hence, an
ensemble of clusters shows a lag before the onset of a first-order decay law.
Ejection of individual charges may take place throughout the lag, division and separation
of the subclusters.

Rate constants for the fission process, derived from the post-lag section of the ensemble's
decay, are shown (plus-symbols) in Fig.~\ref{RateConstants}.  The rate constant for fission
increases much more rapidly with charge than the constant for ejection of individual charges,
and the rate becomes negligible at lower total charges where ejection is still taking place much
faster than neutral evaporation.  Despite the rapid promotion of fission,
it is difficult to pinpoint a threshold
for instability purely on the basis of dynamics, since the rates change smoothly up to the point
where decay by ejection or fission immediately follows the removal of the equilibration
container.

The competition between ejection and fission for $n=309$ and the absence of fission at
smaller $n$ can be understood from an idealised analysis by Consta and Malevanets
of a spherical droplet splitting into two daughter
droplets with conservation of total volume and of total charge \cite{Consta15a}.
Symmetrical decay (fission) is favoured when electrostatic energy dominates over surface
energy, since the decrease in electrostatic energy of separating two highly charged
subclusters compensates for
the large increase in surface area caused by creating two spheres from one with the
same total volume.  This is the regime in our clusters at large $n$.
In contrast, when $n$ is small,
a fractionally significant decrease in electrostatic energy can be achieved by the ejection
of a single charged particle without incurring a large increase in surface energy.
\par
Our procedure for determining the decay rate constants for ejection and fission
relies on being able to
separate trajectories cleanly into portions corresponding to reactants (intact clusters)
and products (separated particles or fragments).  The methods of backdating outbound
ejected particles and monitoring the connectivity of subclusters in fission amount
to selecting reactant--product dividing surfaces in phase space.  Our choices of surface
are not unique but have the advantages of being straightforward to evaluate and,
crucially, not being prone to recrossing by trajectories.  Hence, crossings can be
interpreted decisively as decay events.  In droplets of polar molecules, where decay
may occur via departure of a small cluster \cite{Consta02a}, it should be possible to
adapt the fission criterion to detect division of the droplet into uneven fragments
simply by adjusting the threshold in the number of particles
at which a detached fragment is defined.
\par
To gain more detailed insight into the
fluctuations leading to decay is a more difficult goal that requires construction
of a reaction coordinate capable of faithfully describing the ensemble of decay pathways.
The task has been approached by Consta and coworkers \cite{Consta02a,Consta03a}
for the case of water droplets containing simple ions.  Such clusters develop a
bottleneck as the ions start to separate prior to decay of the droplet, even when
decay is uneven with respect to fragment size.  A special
transfer reaction coordinate was introduced to identify the resulting dumbbell-shaped
configurations, as distinct from more general prolate distortions.
This approach allowed differences to be observed in the diffusive nature of trajectories
near the top of the free energy barrier for clusters containing different
ions \cite{Consta03a}.  As in any activated process, the height and shape of the
free energy barrier for a given process and system depend on the choice of reaction
coordinate, but the overall rate constant should not, provided that a converged
transmission coefficient can be obtained.

\section{Concluding remarks}

Our fluctuating-container simulations effectively cast charge-driven instability as
a transition under well defined equilibrium thermodynamic conditions.  This approach
has the advantages of not requiring any assumptions about the mechanism of the
instability, of permitting comparisons between different clusters if required, and of
being quite sharply defined even for small clusters (where temperature-driven
transitions are normally broadened).  The method is also oblivious to any transient
evaporation of neutral fragments that are of no interest in the context.  The only
parameter that must be fixed is the imposed pressure, but this quantity does have a
direct physical interpretation.
\par
It is not straightforward to compare our results for the stability of the 309-atom
LJ cluster with the prediction of Rayleigh's classic formula, not least
because the surface tension is not known.  Clusters in this size range have a
dramatically depressed melting temperature in comparison with the bulk, and the
liquid-like regime that we have studied around $T^*=0.43$ is well below the
triple point of bulk Lennard-Jones, which lies at about $T^*=0.69$
\cite{Hansen69a,Mastny07a}.  Below this point, no bulk equilibrium
value of $\gamma$ exists.
\par
The first mode of deformation to become unstable
in Rayleigh's model is the second-rank spherical harmonic, corresponding to
oblate--prolate deviations from the sphere.  Although Rayleigh's work makes
no formal prediction about the mechanism or products that result from the instability,
we note that a prolate deformation much more closely resembles the motion
that precedes the fission mechanism that we observe in the LJ cluster at large
number $n$ of charges.  Nevertheless, even where single-particle ejection
dominates the decay and fission is absent,
the clusters are typically distorted in a prolate direction.
\par
For simplicity, we have given all particles in our cluster the same LJ energy
and length parameters.  However, if charged particles interacted with stronger
well depth $u$ or had a larger diameter $\sigma$, one could envisage emission of
``solvated'' charges rather than bare particles.  Clearly, a number of other
refinements could be included to make the model a more realistic depiction of
a particular system.  However, the idealised model has allowed a systematic
survey of charge limits, the decay of large ensembles, rate constants, and activation
barriers to be performed.  The methods deployed here should be useful in
future investigations of other charged clusters and droplets.

\section*{Acknowledgements}

SJG is grateful to Durham University's Institute for Advanced Research Computing for
a summer seedcorn grant.  DAB is indebted to CNRS (Centre National de la Recherche
Scientifique) for the award of a Chaire d'excellence.
Travel associated with this collaboration was supported by the Alliance
Programme of the British Council and the French Minist{\`e}re des Affaires Etrang{\`e}res.
The authors thank Dr Florent Calvo for helpful discussions.

\vspace{12pt}

\end{document}